\documentclass[fleqn,usenatbib]{mnras}
\usepackage{amsmath}
\usepackage{graphicx}
\usepackage{amssymb,txfonts}
\usepackage{float}

\newcommand{\msun}{{\rm M}_{\sun}}
\newcommand{\rsun}{{\rm R}_{\sun}}
\newcommand{\lsun}{{\rm L}_{\sun}}

\newbox\grsign \setbox\grsign=\hbox{$>$} \newdimen\grdimen \grdimen=\ht\grsign
\newbox\simpropbox
\setbox\simpropbox=\hbox{\raise.5ex\hbox{$\propto$}\llap
 {\lower.5ex\hbox{$\sim$}}}\ht2=\grdimen\dp2=0pt

\title[The Sch{\"o}nberg--Chandrasekhar limit]{The nature of the Sch{\"o}nberg--Chandrasekhar limit}

\author[J. Zi\'o{\l}kowski and A. A. Zdziarski]{Janusz Zi\'o{\l}kowski\thanks{E-mail:
jz@camk.edu.pl, aaz@camk.edu.pl} and Andrzej A. Zdziarski{\textcolor{blue}{\footnotemark[1]}}\\
Nicolaus Copernicus Astronomical Center, Polish Academy of Sciences,
Bartycka 18, PL-00-716 Warszawa, Poland}

\begin{document}


\pagerange{\pageref{firstpage}--\pageref{lastpage}} \pubyear{2020}

\maketitle

\label{firstpage}

\begin{abstract}
We present a comprehensive description of the Sch{\"o}nberg--Chandrasekhar (S--C) transition, which is an acceleration of the stellar evolution from the nuclear to the thermal time scales occurring when the fractional mass of the helium core reaches a critical value, about 0.1. It occurs in the 1.4 to 7 $\msun$ mass range due to impossibility of maintaining the thermal equilibrium after the nuclear energy sources in the core disappear. We present the distributions of the hydrogen abundance, the energy generation rate and the temperature for stars crossing that limit. We confirm that a sharp S--C limit is present for strictly isothermal cores, but it is much smoother for real stars. The way the boundary of the core is defined is important for the picture of this transition. With a strict definition of the core as the region where the helium abundance is close to null, it occurs in an extended range of the fractional core mass of roughly 0.03 to 0.11. The cause of that is a gradual core contraction causing a correspondingly gradual loss of the core isothermality with the increasing core mass. On the other hand, when using definitions allowing for some H abundance in the core, the S--C transition is found to be sharper, at the fractional core mass of between about 0.07 and 0.11. Still, it is more a smooth transition than a sharp limit. We have also searched for specific signatures of that transition, and found that it is associated with the stellar radius first decreasing and then increasing again. We have considered whether the S--C limit can be used as a diagnostic constraining the evolutionary status of accreting X-ray binaries, but found such uses unfounded.
\end{abstract}

\section{Introduction}
\label{intro}

The concept of the Sch{\"o}nberg--Chandrasekhar (hereinafter S--C)
limit is one of the classic results of the early stellar evolution
theory. Its history started with the suggestion by \citet{g38} that
a helium core left after the hydrogen exhaustion in the central part
of a star should become isothermal owing to the lack of the energy
sources within the core. Next, \citet{hc41} calculated models
containing such isothermal cores but not consisting of He; they
constructed homogeneous models in which chemical composition was the
same for the core and the envelope. They found that it was possible
to construct a model if the relative mass of the isothermal core,
$q_{\rm c}\equiv M_{\rm c}/M$, was less than about 0.38. In a later
paper, \citet{sc42} constructed more physically realistic models in
which the configuration was composed of a He core and an H-rich
envelope. Such configurations are a natural outcome of the core
H-burning phase. They found the limit of
\begin{equation}
q_{\rm c}\lesssim 0.37\left(\frac{\mu_{\rm env}}{\mu_{\rm c}}\right)^2,
\label{qc}
\end{equation}
which has become known as the S--C limit. Here, $\mu_{\rm env}$ and
$\mu_{\rm c}$ are the mean molecular weights of the envelope and
core, respectively. For $\mu_{\rm env}=0.6$ (about cosmic
composition) and $\mu_{\rm c}=1.3$ (He dominated), $q_{\rm c}\approx
0.08$. Above this core mass, the core collapses. This leads to an
acceleration of the evolution, from a nuclear to a thermal
time-scale. This also represents the onset of the Hertzsprung gap.

Soon, it became clear that the S--C limit cannot be universal.
Models of \citet{sc42} were simplified and, in particular, used the
ideal gas equation of state. As noted by \citet*{hay62}, the He
cores of stars with a total mass of $M\lesssim 1.3\msun$ become
quickly degenerate, and then they remain isothermal even above the
S--C limit. On the other hand, \citet{cg68} claimed that the cores
(defined by the H abundance being relatively low) of stars with
$M\gtrsim 6\msun$ are already heavier than the S--C limit already at
the moment of the central hydrogen exhaustion. Some years later,
\citet{roth73} constructed models of stars after central H
exhaustion. These were idealized in having strictly isothermal
cores, but otherwise included all the relevant physics.
\citet{roth73} constructed evolutionary sequences following the
growth of the core mass owing to H shell burning for several stellar
masses in the 1 to 3 $\msun$ range and confirmed that such idealized
configurations had thermally stable solutions up to certain critical
$q_{\rm c}$ (i.e. up to the S--C limit) and then had to jump
discontinuously to another branch of thermally stable solution with
a much smaller core radius (see, e.g., fig.\ 30.16 of
\citealt*{kip12}). \citet{roth73} also found that the S--C limit
disappears for such idealized stars with $M\lesssim 1.4\msun$, as
well as noted that the real stars would have to evolve through
non-equilibrium models. Indeed, a very fast transition between
stable models near the main sequence and near the giant branch does
occur in the evolution of real stars, forming the Hertzsprung gap.
The PhD thesis results of \citet{roth73} are summarized in detail by
\citet{kip12}.

The S--C limit was later analysed by some other authors, though not
much progress has been achieved with respect to the above results.
\citet{eggl81} and \citet*{eggl98} discussed whether the existence
of the S--C limit could by itself explain the transformation of a
star into a red giant after central hydrogen exhaustion. The answer
was negative; the S--C limit can explain why the core has to contract
but it cannot explain why the envelope has to simultaneously expand.
\citet{beech88} considered configurations similar to those discussed
in the original \citet{hc41} and \citet{sc42} papers.
\citet{beech88} demonstrated that it is possible to derive the S--C
limit by approximating the star with a simplified composite model
with an isothermal core surrounded by an $n=1$ polytropic envelope.
\citet{beech88} found that the limiting relative mass of the
isothermal core is 0.27 for a chemically homogeneous configuration
and 0.10 for the configuration composed of a He core and H-rich
envelope; the latter agrees with that found by \citet{sc42}. The
former differs from the result of \citet{hc41}, having 0.27 instead
of 0.37. This is probably because an $n=1$ polytrope (chosen for
computational convenience) is a worse approximation of the envelope
than the one with a more realistic $n=3$ used by \citet{hc41}.

\citet{eggl98} considered simple analytic two-polytrope models in
which the core and envelope were approximated by $n=5$ and 1
polytropes, respectively. They assumed a jump in the chemical
composition between the core and the envelope. The size of this jump
(namely the ratio, $\alpha$, of the mean molecular weight of the gas
in the core and in the envelope) was a free parameter of the model.
They found that the S--C phenomenon is not present (i.e., the
fractional core mass can be arbitrarily large) for $\alpha<3$. For
$\alpha = 3$, the S--C phenomenon abruptly appears and the value of
the limit is $2/\upi\approx 0.64$, while the classical value is
0.10. The phenomenon is also present for all values of $\alpha>3$.
However, $\alpha \geq 3$ does not correspond to a typical case of a
He core and a Population I envelope, where $\alpha = 2.2$. In the
case of a carbon-oxygen core, $\alpha = 3.3$. Their unexpected
result, namely the lack of the S--C phenomenon in the most typical
case, appears to be due to the use of the $n=5$ polytrope as an
approximation for the core of the star, while the isothermal core
formally corresponds to a polytrope with $n \to \infty$. Indeed,
when using $n > 5$, \citet{eggl98} found the S--C phenomenon at any
$\alpha$.

We note that at present we have tools to construct precise stellar
models, and there is no longer a need to limit the study to models
using polytropes and/or assuming the core to be isothermal. Here, we
use an evolutionary code developed from that of
\citet{paczynski69,paczynski70} to study in detail the behaviour of
stars in the 1 to 7 $\msun$ mass range from the moment of their
leaving the main sequence up to the moving to the giant branch. We
search for signatures of the S--C limit, in particular the
acceleration of the evolution. An issue we find to be crucial is the
way the He core, and thus $q_{\rm c}$, are defined. A possible
option is to define the core as the region where He completely
dominates the composition and the burning of hydrogen is negligible.
This corresponds to the fractional H abundance of $X\lesssim
10^{-6}$. Such a definition is adopted, in particular, in our
stellar evolution code. On the other hand, the S--C limit has been
derived based on the pressure balance at a boundary between the core
and the envelope. Obviously, such boundary has to have a H abundance
intermediate between that of the envelope and of the hydrogen-less
core interior. We have found that $q_{\rm c}$ corresponding to a
given evolutionary phase strongly depends on the definition of
$q_{\rm c}$. If the strict definition (no hydrogen in the core) is
adopted, the acceleration already starts at $q_{\rm c}\approx 0.03$
and the core mass gradually increases while going through the formal
S--C limit. With that strict definition, there is no S--C limit, but
rather an extended S--C transition, with the core losing
isothermality and collapsing over a relatively wide range of $q_{\rm
c}$. On the other hand, if a less strict (but still physically
justifiable) definition is adopted, then the S--C transition is sharper. Here, we use two such less strict definitions, namely $q_{\rm c1}$ corresponding to the radius at which the core switches from contracting to expanding, and
$q_{\rm c2}$ corresponding to the radius where $X=0.01$ is reached.

In our study, we also attempt to determine whether the S--C limit
can be used to constrain the evolutionary status of a star in the
absence of detailed spectroscopic information, especially in the
case of donors of accreting X-ray binaries, in which case some
specific S--C constraints were applied in the past.

In order to clarify these uncertainties about the meaning and nature
of the S--C limit, we have constructed detailed evolutionary models
of stars crossing that limit. We have studied the related
evolutionary effects, and attempted to provide a detailed
description of of the evolution, in particular with respect to
evolutionary time and the core mass, and paying attention to the way
the core is defined. We have also searched for any additional
signatures of this crossing.

\begin{figure}
\centerline{\includegraphics[width=\columnwidth]{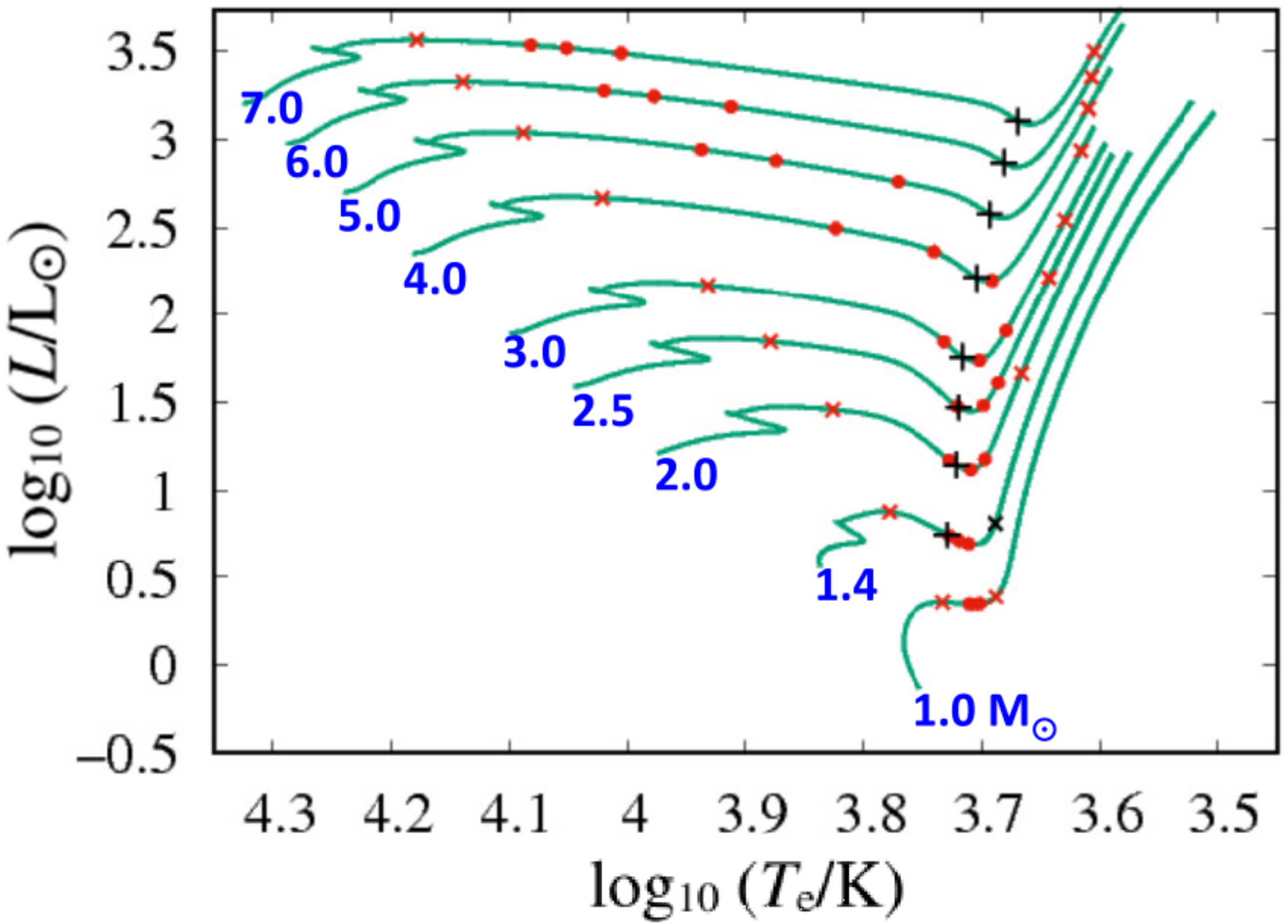}}
 \caption{Evolutionary tracks in the H--R diagram (the luminosity vs.\ the effective surface temperature) for stars in the mass range of 1 to 7 $\msun$. The masses are marked at the beginning of each track. Red crosses ($\times$) show the points where $q_{\rm c0}=0.06$ (left) and 0.12 (right). The red dots show the points where $q_{\rm c0}=0.09$, 0.095 and 0.1 (from left to right). The black signs (+) correspond to the location of the kinks of the stellar radius as a function of time, shown in Fig.\ \ref{Rt} below. The kinks visible to the left of those discussed above correspond to the exhaustion of H at the stellar centre, when the star leaves the main sequence.}
 \label{hrd}
 \end{figure}

\section{The evolutionary models}
\label{models}

We use the Warsaw stellar evolution code of
\citet{paczynski69,paczynski70}, then developed by M. Koz{\l}owski
and R. Sienkiewicz. Its main current features and updates (e.g., the
opacities, nuclear reaction rates and equation of state) are
described by \citet{pamyatnykh98} and \citet{ziolkowski05}.
\citet{z16}, calibrated the code to reproduce the Sun at the solar
age. This resulted in a H mass fraction of $X=0.74$, a metallicity
of $Z=0.014$, and a mixing length parameter of 1.55 (adopting the version of the mixing theory of \citealt{paczynski69}). Here, we
use these values to follow evolution of stars in the mass
range\footnote{We have considered the evolution of an $8\msun$ star,
but for this mass the central He burning already takes place during
the relevant evolutionary phase, and our numerical code is not
accurate in that regime. Thus, the results presented in this work
concern the mass range of 1 to 7 $\msun$.} of 1 to 7 $\msun$ up to
the advanced (with a luminosity of more than $10^3 \lsun$) red giant branch.

The evolutionary tracks in the Hertzsprung--Russell (H--R) diagram are shown in Fig.~\ \ref{hrd}. The transition through the S--C phase happens after hydrogen is exhausted at the stellar centre. This exhaustion corresponds to the second direction reversal (counting from the left-hand side) on the shown evolutionary tracks for $M\geq 1.4\msun$. At this point, the fractional core mass is null according to its strict definition, with an H abundance of $X\leq 10^{-6}$, which corresponds to its burning being completely negligible. Hereinafter, we denote the fractional mass of the core defined in this way by $q_{\rm c0}$. Then, the minima of $L$ correspond to the Hayashi line, at the transition from the subgiant to the giant branch, and they correspond to $q_{\rm c0}\approx 0.11$.

\begin{figure}
\centerline{\includegraphics[width=1.05\columnwidth]{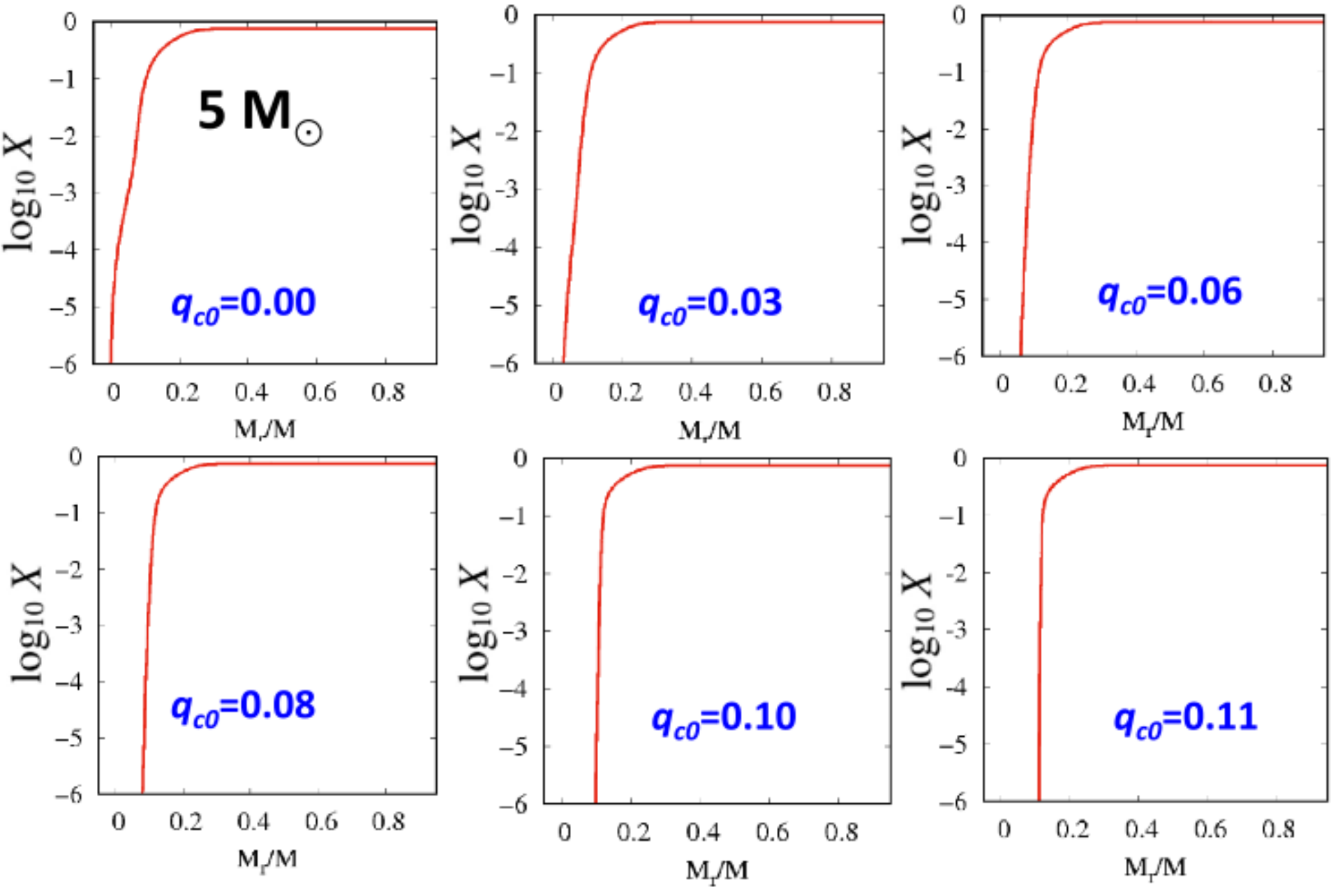}}
 \caption{The H abundance vs.\ the ratio of the enclosed mass for a $5\msun$ star at 6 values of $q_{\rm c0}$. We see that $X$ increases gradually. This effect is especially pronounced for low $q_{\rm c0}$. After the S--C transition, the core boundary becomes well defined.}
 \label{XMr}
 \end{figure}

\begin{figure*}
\centerline{\includegraphics[width=13cm]{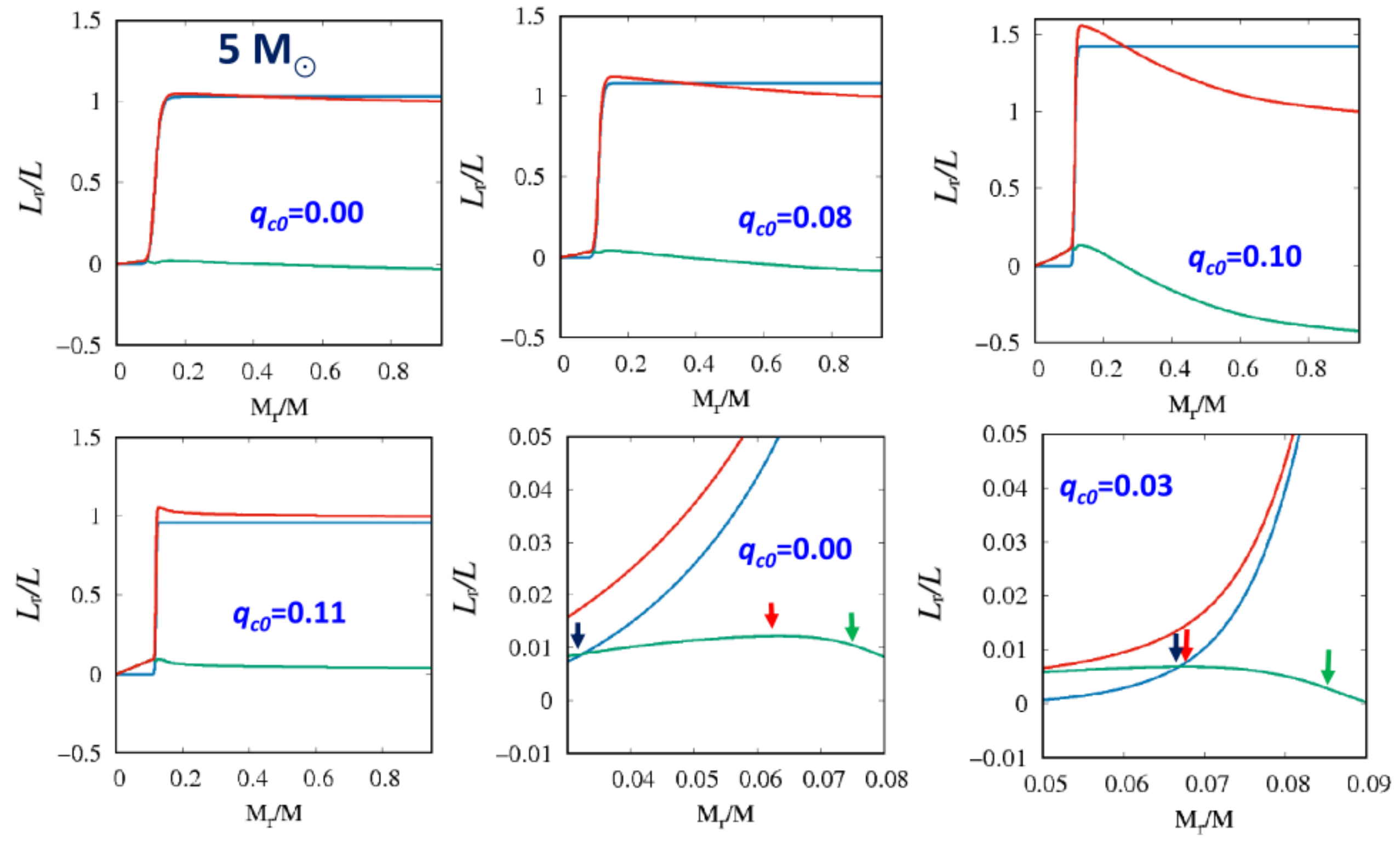}}
 \caption{The local luminosity (i.e., the total energy flux through the surface of a sphere of a given radius), $L_r$, vs.\ the corresponding enclosed mass for a $5 \msun$ star and for different values of $q_{\rm c0}$. We plot $L_r/L$, where $L$ is the total surface luminosity of the star, in terms of $M_r/M$. The green curves show the gravitational energy flux (positive when generated by core contraction and negative by envelope expansion). The blue curves show the flux generated by the nuclear reactions (H shell burning). The red curves shows the sum. We see that after the central H exhaustion, the star gradually departs from the thermal equilibrium and later returns to it when approaching the giant branch. The last two panels zoom to the area near the core boundary for the cases of $q_{\rm c0}=0.00$ and 0.03. The dark blue arrows show the location where the nuclear and gravitational energy fluxes are equal. The red arrows indicate the surface separating the contracting inner part of the core from the expanding outer layers. The green arrows indicate the surface with $X= 0.01$.
}
 \label{LMr}
 \end{figure*}

As we have mentioned above, the H abundance increases with the
enclosed mass rather gradually, especially before crossing the S--C
phase. This is illustrated in Fig.\ \ref{XMr}, which shows the
dependence of $X$ on the (running) mass enclosed within a given radius, $M_r$, for a star with $5\msun$ at 6 values of $q_{\rm c0}$. Thus, we see that the
actual definition of the core boundary is quite fuzzy. Therefore, we
define here two additional fractions. One is $q_{\rm c1}$,
corresponding to the radius at which the core switches from
contracting to expanding. The other is $q_{\rm c2}$, corresponding
to the radius where $X=0.01$. Another possible definition is the
radius at which the energy flux generated in the nuclear reactions
equals the gravitational energy flux (generated by core contraction,
see Fig.\ \ref{LMr}). The $M_{\rm c}/M$ defined that way is
generally close to $q_{\rm c1}$ (except when $q_{\rm c0}= 0$).

Of course, the physical picture of stellar evolution of the star
does not depend on the definition of $q_{\rm c}$. However, as we
show below, the picture of the S--C transition does depend on it.
Namely, the rate of the stellar evolution as a function of $q_{\rm
c}$ during that transition is quite different for different
definitions, and then the transition covers different ranges of
$q_{\rm c}$.

 \begin{figure}
\centerline{\includegraphics[width=1.05\columnwidth]{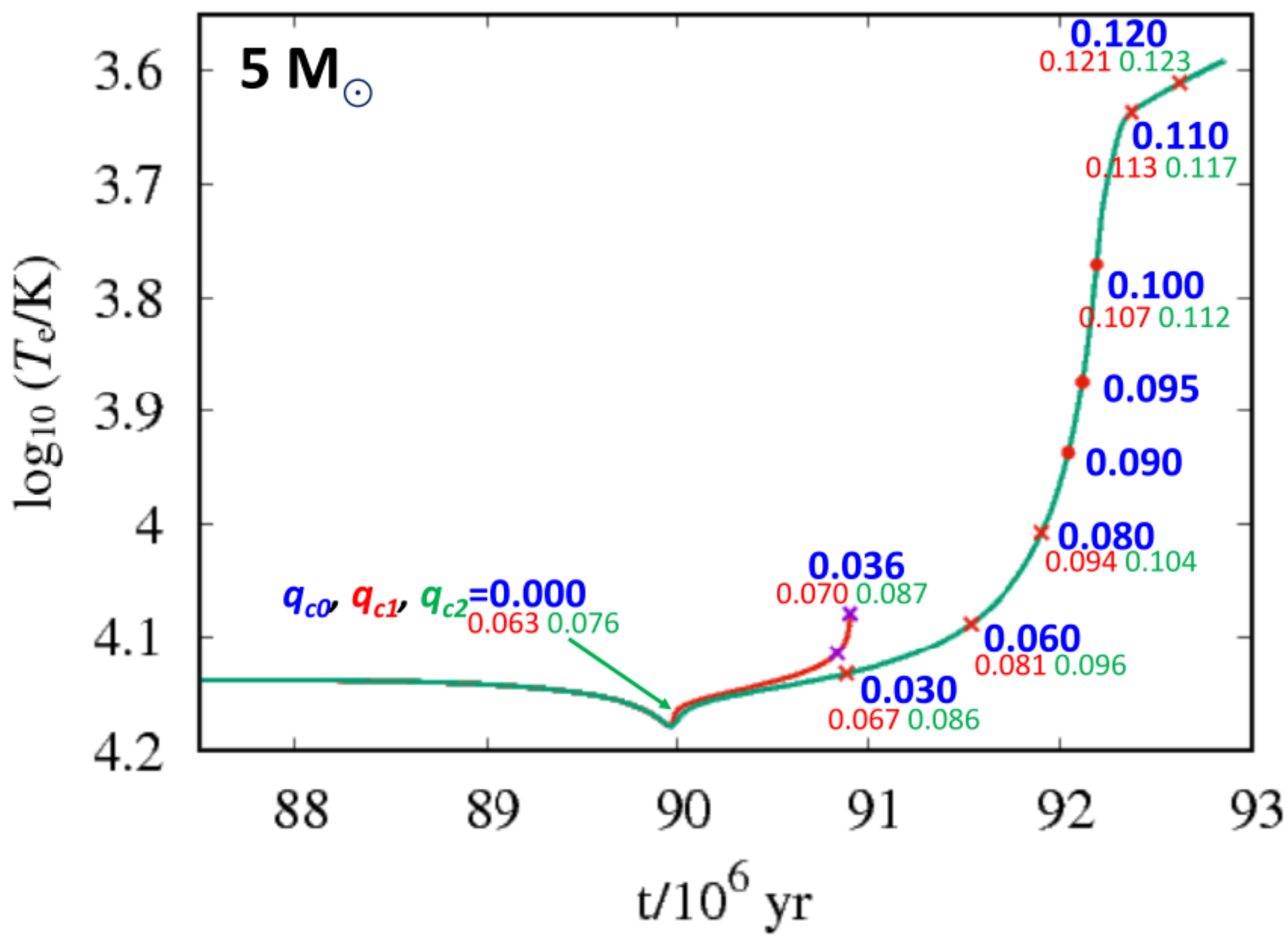}}
 \caption{The green curve shows the stellar effective temperature vs.\ the evolutionary time (from the zero-age main sequence) for a $5\msun$ star. The values of $q_{\rm c0}$ (negligible H), $q_{\rm c1}$ (at the radius at which the core ceases to collapse) and $q_{\rm c2}$ ($X\leq 0.01$) are shown along the track. The left edge of the plot corresponds to the first turn on the H--R diagram, at which the central H content is about 0.05. The point with $q_{\rm c0} = 0$ corresponds to the second turn on the H--R diagram and the time of the central H exhaustion. For comparison, the red curve shows a case when the strict thermal equilibrium is enforced (no gravitational energy release and isothermality of the core). This sequence ends at $q_{\rm c1}\approx 0.070$, $q_{\rm c2} \approx 0.087$. While these correspond to the classical S=-C limit, this case is artificial.}
 \label{Tet}
 \end{figure}

In order to gain further insights into the structure of the core, we
show in Fig.\ \ref{LMr} the local luminosity (i.e., the total energy
flux through the surface of a sphere of a given radius), $L_r/L$, as
a function of the fractional mass enclosed, $M_r/M$, for a $5\msun$
star at different fractional masses of the core defined by $q_{\rm
c0}$. Here $L$ is the total surface luminosity of the star. The
green curve shows the gravitational energy flux (which is greater
than 0 within the contracting core and less than 0 during
expansion). The blue curve shows the flux generated in the nuclear
reactions. The red curve gives the net energy flux, being the sum of
the previous two. The steeply rising parts of $L_r$ on the first
four panels correspond to the H shell burning, which generates most
of the total energy. The energy generated by the collapsing core is
shown by the rising part of $L_r$ below the H-burning shell. We see
that the shell remains around $M_r/M\approx 0.1$ for $q_{\rm c0}$ in
the 0.03--0.10 range, i.e., the mass within the shell can be much
higher than that of the core defined by the completely negligible H
burning. The declining part of $L_r$ corresponds to the expanding
envelope. The envelope is near thermal equilibrium at the beginning
of the S--C transition, at $q_{\rm c0}\lesssim 0.03$, and after it,
at $q_{\rm c0}\gtrsim 0.11$, when it approaches the giant branch.
However, the star is out of thermal equilibrium during that
transition.

The last two panels in Fig.\ \ref{LMr} zoom into the range around
the core boundary for $q_{\rm c0} =0.00$ and $q_{\rm c0} =0.03$. The
arrows show the locations of the core boundary according to the
different definitions of $q_{\rm c}$. In our opinion, the most
physically justifiable definition is that corresponding to the
location of the surface separating the contracting inner part of the
core from the expanding outer layers, i.e., $q_{\rm c1}$.

We then show in Fig.\ \ref{Tet} the time dependence of the effective
temperature, $T_{\rm e}$, around the S--C transition for a $5\msun$
star. At the time of the central H exhaustion, $q_{\rm c0}=0$, but
there is already a He-dominated core with $X\ll 1$, and $X=0.01$ is
reached only at $M_{\rm r}/M=0.076$. At $q_{\rm c0}=0.03$, $q_{\rm
c}$ defined by the transition from collapse to expansion and by
$X=0.01$ is already about twice and thrice as high, respectively.
The acceleration of the evolution of the effective temperature
starts around this point, and the rate of the decrease of $T_{\rm
e}$ keeps increasing up to $q_{\rm c0}\approx 0.11$, which
corresponds to the transition to the giant branch. Thus, the S--C
phenomenon can be described as a quite extended phase in $q_{\rm
c0}$, from 0.03 to 0.11, but it is much more rapid in terms of
$q_{\rm c1}$, or even more so of $q_{\rm c2}$, from about 0.09 to
0.12. Still, the phenomenon is more a smooth transition than a sharp
limit.

The red curve in Fig.\ \ref{Tet} shows the behaviour of a $5\msun$
configuration for which strict thermal equilibrium is enforced (no
gravitational energy release or consumption and strict isothermality
of the core). This sequence ends at the fractional core mass
corresponding to $q_{\rm c0}\approx 0.036$, $q_{\rm c1}\approx
0.070$ and $q_{\rm c2} \approx 0.087$. This case illustrates the
classical assumption used to derive the S--C limit, where there are
models satisfying this assumption only below the limit. However,
real stars do not behave this way, and follow the sequence shown by
the green curve. Along it, the process of the departure from the
thermal equilibrium initially accelerates then later decelerates but
is always gradual. It is then difficult to single out a specific
fractional core mass and define it as the S--C limit.

\begin{figure*}
\centerline{\includegraphics[width=\textwidth]{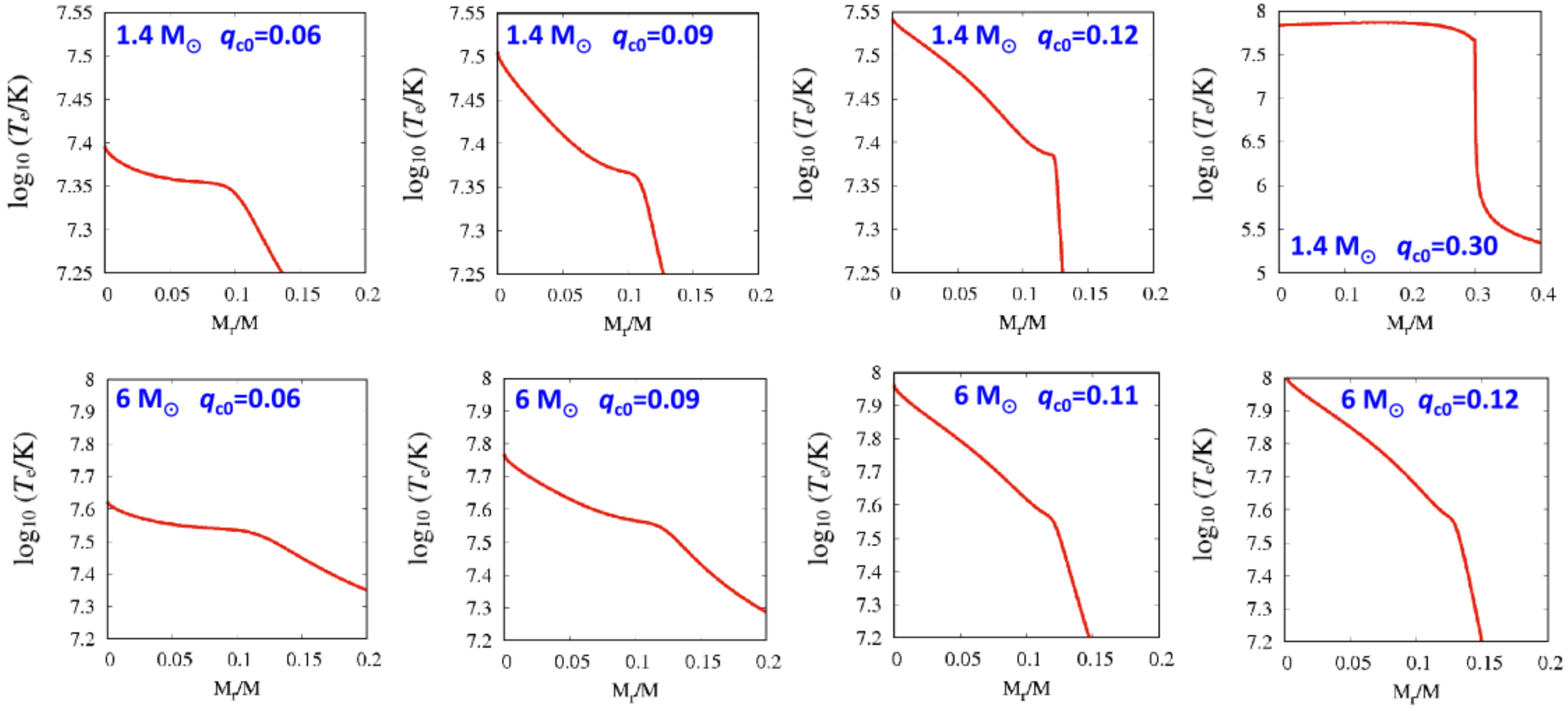}}
 \caption{The interior temperature as functions of the mass enclosed for $M=1.4$ and $6 \msun$ and for different fractional core masses (with its strictest definition $q_{\rm c0}$). We see the core loses isothermality (and thus starts to collapse) relatively quickly, at $q_{\rm c0}\gtrsim 0.06$. At the expected S--C limit, $q_{\rm c0}= 0.1$, it is already far from isothermal. On the other hand, the core becomes isothermal again for $1.4 \msun$ and $q_{\rm c0} =0.30$ because electrons become degenerate.}
 \label{TMr}
\end{figure*}

Fig.\ \ref{TMr} shows the distributions of the interior temperature
as functions of the mass enclosed for stars of $M=1.4$ and $6 \msun$
and for different $q_{\rm c0}$. We see that the He cores lose their
isothermality and thus start to contract soon after the central
hydrogen exhaustion. Initially this contraction is relatively slow.
For $q_{\rm c0} = 0.03$, the core remains approximately isothermal.
Then, the contraction accelerates, and when the core approaches the
expected S--C limit ($q_{\rm c0}\approx 0.1$), it is already very far
from being isothermal. Thus, the basic assumption used in deriving
the S--C limit is not satisfied. We also see the expected return to
isothermality for larger cores at $1.4\msun$ because the core
becomes degenerate.

\begin{figure*}
\centerline{\includegraphics[width=\textwidth]{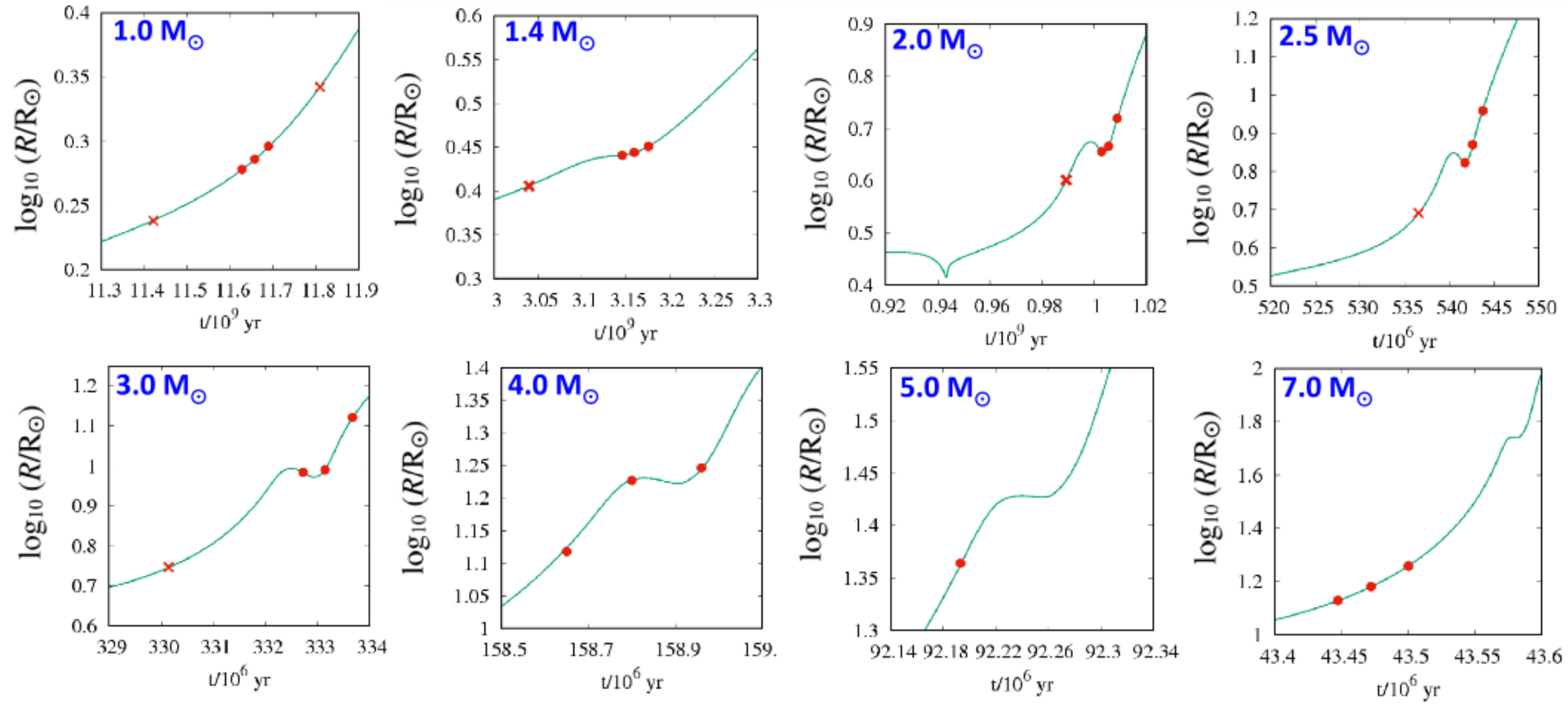}}
 \caption{The stellar radius vs.\ the evolutionary time (from the zero-age main sequence) for $1 \le M/\msun \le 7$ stars. The red dots show the points where $q_{\rm c0}=0.09$, 0.095 and 0.1 (from left to right), and the red crosses show the points where $q_{\rm c0}=0.06$ (left) and 0.12 (right). The single red dot for $M=5\msun$ star corresponds to $q_{\rm c0}= 0.1$. The kinks in the shape of $R(t)$, i.e., at ${\rm d}R/{\rm d}t=0$, occur close to $q_{\rm c0}= 0.1$. They are present for $1.4 \le M/\msun \le 7$. The lower kink in $R(t)$ for $2\msun$ corresponds to the time of the central hydrogen exhaustion, $q_{\rm c0} = 0$. }
 \label{Rt}
 \end{figure*}

We have also found an interesting and unexpected behaviour of the
radius of the star, namely first a local decrease then a local
increase, see Fig.\ \ref{Rt}. We have found this phenomenon always at the fractional mass of the He core of $q_{\rm c0}\approx 0.1$, i.e., around the end
of the S--C transition, and for the mass range of 1.4 to 7 $\msun$,
where this phenomenon takes place. The black crosses in Fig.\
\ref{hrd} mark the points at which the radius as a function of time,
$R(t)$, executes the wiggle shown in Fig.\ \ref{Rt}, starting to increase again
after a temporary decrease.

The values of $q_{\rm c0}$ for which this happens increase rather
slowly with $M$, 0.090 for $M=1.4\msun$, 0.098 for $4\msun$, 0.109
for $6\msun$ and 0.114 for $7\msun$, i.e., in the narrow range of
$q_{\rm c0}$ of 0.09--0.11. This narrowness is surprising, taking
into account large spans of other parameters of the stars at this
phase of the evolution. For example, we have the radius of $R=2.76
\rsun$ and the central temperature of $3.0\times 10^7$\,K for
$M=1.4\msun$ and $54.8 \rsun$, $1.0\times 10^8$\,K for $7\msun$.
Thus, this phenomenon appears to be caused in some way by the core
collapse, but details remain unclear. In Fig.\ \ref{hrd}, we also
see that the kink in $R(t)$ approximately coincides with the star
moving from the subgiant branch of into red giant branch. (We use
here the terms subgiant and red giant in the sense of the location
in the H--R diagram.) We note that changes in the character of the
time dependence  occur only for the stellar radius. The luminosity
evolution does not show any features, and it monotonically decreases
through the evolutionary phase of interest.

Coming back to Fig.\ \ref{Tet}, we see that the rate of the
evolution (as measured by the changing effective temperature; the
changes of $L$ are even slower, see Fig.\ \ref{hrd}) after the
central H exhaustion remains rather slow for some time while the
star has already quite an extended core composed mostly of helium.
In the considered case of $5\msun$, there is a zone of avoidance
(the Hertzsprung gap) for $3.7 \lesssim \log_{10}(T_{\rm e}/{\rm K})\lesssim 4.0$, but subgiants with higher $T_{\rm e}$ move through that phase
relatively slowly. An analogous situation, with a somewhat different
range of the gap, occurs for other stellar masses. Thus, evolved
stars are common on both sides of the gap.

On the other hand, \citet{king93} placed a limit on the mass of the
donor in the binary V404 Cygni by arguing that the mass of its core
has to be above the S--C limit because otherwise the star would not
have left the main sequence. However, the S--C limit occurs quite
some time after leaving the main sequence. One could rather talk
about the transition from a subgiant to giant, or from the
luminosity class IV to III. The argument of \citet{king93} was then
used by \citet*{munoz08} to constrain the donor mass in the binary
GX 339--4. Subsequently, \citet{heida17} used the constraint of
\citet{munoz08} to place a strict upper limit on the mass of the
black hole and a lower limit on the inclination of that binary.
These constraints have then been later widely used in papers on that
binary, see a discussion in \citet*{zzm19}. However, the two best-fitting stellar templates found by \citet{heida17} were one of the luminosity class III (a giant) and one of the class IV (a subgiant). Thus, evolved stars on both sides of the S--C limit were found to fit their observational data.
Furthermore, the claimed maximum donor masses for V404 Cyg and GX
339--4 are about $1.3\msun$ and about $1.1\msun$, respectively,
where the S--C limit does not apply. We also note that the numerical
value of the limit used by \citet{king93}, 0.17, is significantly
above the canonical value of about 0.10.

\section{Conclusions}

We have studied in detail the evolution of the stars in the 1 to 7
$\msun$ mass range from the time of leaving the main sequence up to
reaching the giant branch, with the goal of presenting a
comprehensive description of the Sch{\"o}nberg--Chandrasekhar
transition. This transition occurs in the 1.4 to 7 $\msun$ mass
range. An important issue, which has apparently received no
attention in the literature, is the way the He core is defined. When
using the strict definition of the core as the region with the He
abundance is close to null (with the fractional core mass denoted as
$q_{\rm c0}$) the S--C transition does not correspond to a sharp
limit but instead occurs in an extended range of the fractional core
mass, of 0.03 to 0.11. The cause of this is a very gradual core
contraction causing a correspondingly gradual loss of the core
isothermality with the increasing core mass. On the other hand, when
using less strict definitions allowing for some H abundance, e.g.,
$q_{\rm c1}$ defined by the radius at which the core stops its
collapse, or $q_{\rm c2}$ defined by $X\leq 0.01$, the S--C
transition is found to be sharper, at the fractional core mass of
about 0.07 to 0.11 and 0.09 to 0.12 for the former and latter definition, respectively. Thus, we should call this phenomenon an S--C transition rather than a limit. Therefore, it is difficult to single out a fractional core
mass and define it as the S--C limit.

In Fig.\ \ref{XMr}, we illustrated the problem of a precise
definition of the core boundary by showing the distribution of the H
abundance vs.\ the mass within a given radius. This figure shows
that the H abundance increases very gradually at the beginning of
the S--C transition. This increase becomes much sharper when the
transition is completed. Similarly, the power generated within a
given mass increases gradually at low core masses, as shown in Fig.\
\ref{LMr}. Fig.\ \ref{Tet} then shows an example of the temporal
evolution of the effective temperature, showing the fractional core
mass according to our three adopted definitions. It illustrates our
statement that the acceleration of the temporal evolution occurs
very gradually when the strict core definition, $q_{\rm c0}$, is
adopted, while the acceleration is occurs much faster for the other
two definitions. Fig.\ \ref{TMr} shows that the interior temperature
decreases gradually even before the beginning of the S--C transition
(whereas isothermality was the basic assumption made in the original
derivation of the S--C limit).

For comparison, Fig.\ \ref{Tet} also shows the behaviour of a
$5\msun$ configuration for which a strict thermal equilibrium was
enforced (no gravitational energy release or consumption and a
strictly isothermal core). In this case, in which the assumptions
made in the derivation of the S--C limit are fully satisfied, we
indeed find unambiguously the existence of a sharp S--C limit.
Namely, we obtain models before reaching this limit, for which the
fractional core mass is 0.070 or 0.087, depending on the definition,
while there are no models satisfying our adopted assumptions above
the limit. However, this case does not correspond to real stars.

We have also searched for a specific signature of the transit of the
S--C limit, and found one. Namely, the stellar radius shows a wiggle
at that transition, first decreasing and then increasing, as shown
in Fig.\ \ref{Rt}.

We have also considered whether the S--C limit can be used as a
diagnostic to constrain the evolutionary status of accreting X-ray
binaries, but we have found such uses to be unfounded. The S--C
limit does approximately correspond to the boundary between the
subgiant and giant branches in the mass range of 1.4 to 7 $\msun$.
However, in the specific cases when it was used in the past, the
giant/subgiant distinction could not be made observationally and the
claimed limiting stellar masses were less than 1.4 $\msun$.

\section*{Acknowledgements}
We thank the anonymous first referee for a valuable report on the original
version of this work, and the second referee, Prof.\ Christopher Tout, for a
very careful analysis of the revised version. This research has been
supported in part by the Polish National Science Centre under the
grants 2015/18/A/ST9/00746 and 2019/35/B/ST9/03944.

\section*{Data Availability}

There are no new data associated with this article.

\label{lastpage}
\end{document}